\documentclass[a4paper,cits]{PoS}

\def\apj{ApJ}%
\def\apjl{ApJ}%
\def\aap{A\&A}%
\def\aapr{A\&A~Rev.}%
\def\mnras{MNRAS}%
\def\prd{Phys.~Rev.~D}%
\def\nat{Nature}%
\title{X-ray reverberation in NLS1}

\ShortTitle{X-ray reverberation in NLS1}

\author{{\speaker{Lance~Miller}$^{a}$ and T.~Jane~Turner$^{b}$}\\
        \llap{$^a$}Dept. of Physics, Oxford University, U.K.\\
        E-mail: \email{L.Miller@physics.ox.ac.uk}\\
        \llap{$^b$}Dept. of Physics, University of Maryland Baltimore County, U.S.A.\\
        E-mail: \email{tjturner@umbc.edu}}

\abstract{
Reverberation from scattering material around the black hole in active galactic
nuclei is expected to produce a characteristic signature in a Fourier analysis
of the time delays between directly-viewed continuum emission and the scattered light.
Narrow-line Seyfert\,1 galaxies (NLS1) are highly variable at X-ray energies, and are
ideal candidates for the detection of X-ray reverberation.  We show new analysis
of a small sample of NLS1 that clearly shows the expected time-delay signature,
providing strong evidence for the existence of a high covering fraction of
scattering and absorbing material a few tens to hundreds of gravitational radii
from the black hole.  We also show that an alternative interpretation of time delays
in the NLS1 1H\,0707--495,
as arising about one gravitational radius from the black hole, is strongly disfavoured 
in an analysis of the energy-dependence of the time delays.
}

\FullConference{Narrow-Line Seyfert 1 Galaxies and their place in the Universe - NLS1,\\
		April 04-06, 2011\\
		Milan Italy}

\begin{document}

\section{Introduction}
The measurement of reverberation at optical wavelengths has become a powerful technique
for estimating the size of the broad-line region in active galactic nuclei (AGN), and
combining with line velocity widths allows estimates of black hole mass in AGN to
be made (e.g. \cite{peterson}).  Here we describe the characteristic signature we
expect from reverberation and show evidence that this signature has been detected
at X-ray energies, implying the existence of large global covering factors of
material a few light hours from the black hole, corresponding to a few tens to
hundreds of gravitational radii (r$_g \equiv GM/c^2$).

\section{Reverberation signatures in Fourier space}
\subsection{The analysis of data}
Reverberation signatures are measured by cross-correlating two time series and searching
for time delays between them.  In AGN at optical wavelengths, one of the time series is
the flux measured in a continuum bandpass, the other the flux measured in
an emission line.  A measured time delay between these two gives us the light travel
time between the continuum source and the line-emission region, averaged over the
distribution of circumnuclear material.  

A natural approach to measuring cross-correlation time delays is to work in Fourier
space, and we shall see below that reverberation creates a characteristic signature
in the Fourier phases as a function of the frequency of variation of the source.  
However, we cannot simply take the Fourier transforms of the two light curves to make
such an analysis, for several reasons:
\begin{enumerate}
\setlength{\itemsep}{-1ex}
\item Observed time series are discretely sampled, with non-uniform coverage and large
gaps between multiple observations.  If Fourier transformed, the window function
of the observations would completely dominate the signal.  
\item Observed time series have measurement noise, which adds a floor to the 
powerspectrum and which may bias the measured time delay if uncorrected.
\item Any one set of data is just a particular realisation of the source's variations:
it is a snapshot in time.  In order to correctly estimate the uncertainties in our
measurements, we must allow for the expected statistical variations in the source.
\end{enumerate}
To tackle these issues, we have developed a maximum-likelihood approach, which finds
the powerspectra and cross-powerspectra that best fit the time-domain data, taking
full account of the sampling and noise, and with estimated measurement uncertainties
that fully account for the ``cosmic variance'' of item (3) above.  
The method is based on the ``direct likelihood'' approach used for analysis of 
small cosmic microwave background datasets \cite{bond}.
More details are given by Miller et al. \cite{miller10a,miller10b}.

\subsection{The reverberation signature}
In the following work, we shall consider the ``lag spectrum'': 
the time lags between two time series 
as a function of the frequency of the source variation.  These time lags $\tau$ are
obtained from the phases $\phi$ of the Fourier transform of the cross-correlation
function, $\tau = \phi/\omega$, where $\omega$ is the angular frequency.

\begin{figure}
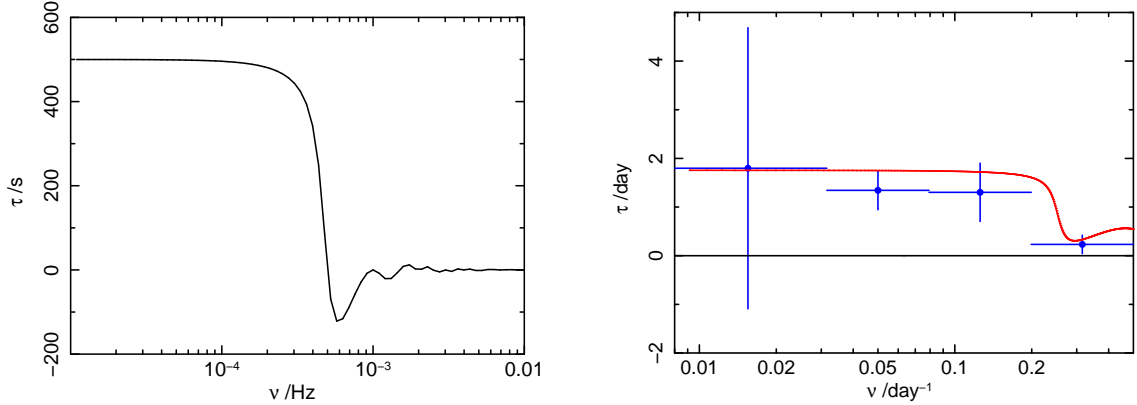

\begin{center}
\resizebox{0.48\textwidth}{!}{
\rotatebox{-90}{
\includegraphics{lagsillustration.ps}
}}
\hspace*{5mm}
\raisebox{-2.2mm}{
\resizebox{0.462\textwidth}{!}{
\rotatebox{-90}{
\includegraphics{lags_with_model.ps}
}}}
\end{center}
\caption{\emph{Left panel:} The lag spectrum expected from a simple partial-covering
thin shell of diameter 2000\,light-seconds, 
showing the rapid transition in time delay at $\nu \simeq 5\times 10^{-4}$\,Hz,
when the time period of the Fourier mode
equals the light travel time across the shell (the low-frequency time delay of 500\,s
has been diluted
in this realisation by adding direct light into the scattered-light time series).
\emph{Right panel:} The lag spectrum for the H$\beta$ and optical continuum 
variations in NGC\,4051 observed by \cite{denney09}, measured here using the
maximum-likelihood method.  The data (blue points with
error bars) show a
rapid transition at $\nu \simeq 0.2$\,day$^{-1}$, 
indicating the distance across the emission-line region to
be about 5 light days, with a mean lag of about 1.8 days.  
The red curve shows an illustrative model (not a best-fit) of an
inclined scattering annulus.
\label{fig1}
}
\end{figure}

Consider a simplified case of a thin spherical shell of material surrounding an AGN, where
directly-seen light is emitted from a point-like source at the centre and measured in
one time series, and light scattered from the shell is observed in a second time
series.  We get some interesting effects if we make the shell ``partial-covering''
by removing the parts of the shell that are both closest to us and furthest away,
so that the shell resembles a ring seen end-on.  Fig.\,\ref{fig1} shows the
expected lag spectrum for this case.  At high frequencies, where the mode time period 
is much shorter than the light travel time across the shell, scattering from different
parts of the shell add incoherently, with a wide range of phases, so the net time
delay tends to zero, with some oscillations that depend on the shell structure.  
At low frequencies, where the period is much longer than the light travel time across
the shell, the scattered light adds mostly coherently and we see a mean time delay
for the shell.  At a frequency where the time period equals the light travel time
across the shell, we see a sharp transition between these two regimes, with the
time delay even plunging negative for some ranges of frequency.  Of course, all the
scattered light signals actually are delayed:  the negative lag behaviour arises
because we are looking in Fourier space, and at this frequency the phases are starting
to wrap around by more than $2\pi$ radians.  If the shell were symmetric and full-covering, the time delays oscillate but stay positive: lags can change sign in the case of 
a partial-covering shell as shown here.

The behaviour from this simple model is broadly reproduced by a wide range of more
realistic scattering scenarios with scattering from a range of radii \cite{miller10a}:  
a mean lag at low frequencies, transitioning to
oscillatory and small time lags at high frequency.  The detailed oscillatory structure
tells us about the reverberation structure and geometry, but the basic signature that
we should look for is the dramatic change in lag at some characteristic frequency,
the value of which tells us the size of the scattering region.

\subsection{Test on optical data}
We can test the method on the optical dataset of NGC\,4051 \cite{denney09}.  
Fig.\,\ref{fig1} shows the lag spectrum, showing the expected rapid transition, 
indicating a total
light travel time across the emission line region 
(from the nearest to the furthest scattering region)
of about 5 days, with a mean lag
of about 1.8 days, consistent with the previous analysis of this data, but showing here
the characteristic frequency-dependent behaviour.

\section{X-ray reverberation detected in narrow-line Seyfert\,1s}
Our goal here is to attempt to detect reverberation in AGN at X-ray energies,
using the maximum-likelihood to take full account of the sampling and noise.
As we are then observing generally much more highly ionised material, and as 
many AGN show rapid (ks) X-ray variability, we may be able to probe the 
environments around the black hole on scales of light-hours rather than light-days.

\begin{figure}
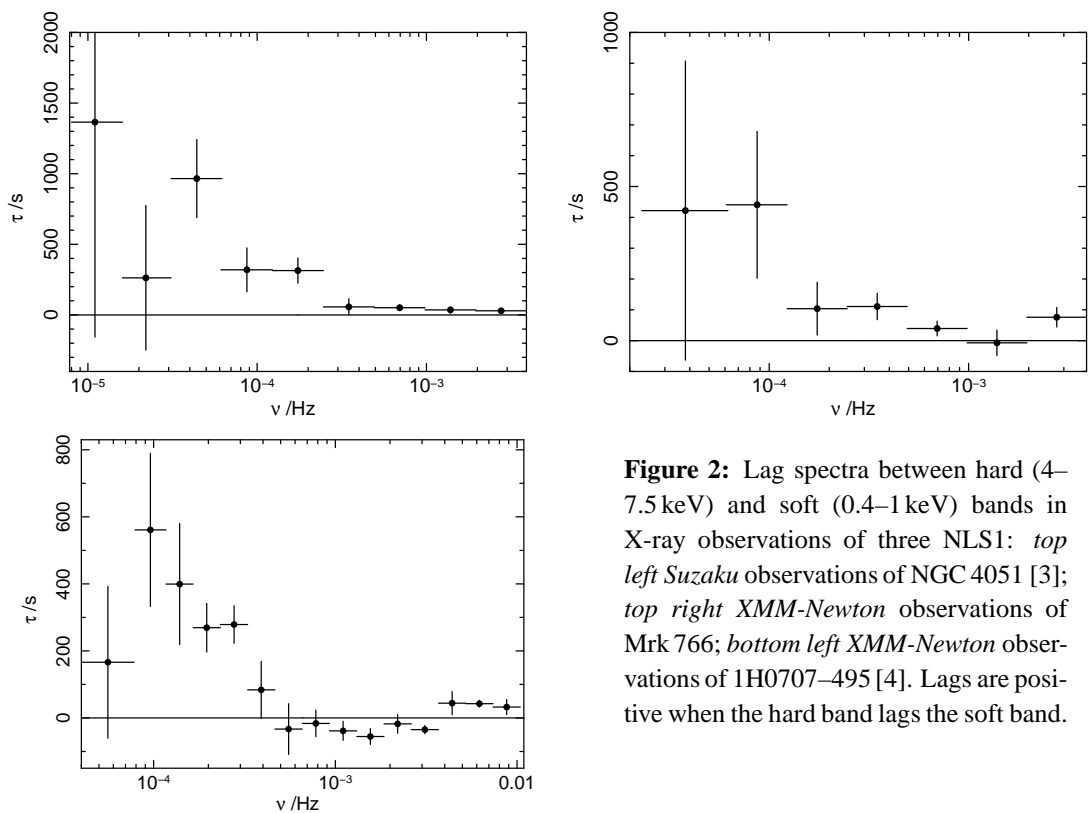

\begin{center}
\resizebox{0.46\textwidth}{!}{
\rotatebox{-90}{
\includegraphics{ngc4051lags.ps}
}}
\hspace*{3mm}
\resizebox{0.46\textwidth}{!}{
\rotatebox{-90}{
\includegraphics{mkn766lags.ps}
}}

\resizebox{0.46\textwidth}{!}{
\rotatebox{-90}{
\includegraphics{1h707softhardnewerrors.ps}
}}
\hspace*{8mm}
\raisebox{-2.4cm}{
\begin{minipage}{0.39\textwidth}{
\caption{
Lag spectra between hard (4--7.5\,keV) and soft
(0.4--1\,keV) bands in X-ray observations of three NLS1: 
\emph{top left} \emph{Suzaku} observations of NGC\,4051 \cite{miller10a}; 
\emph{top right} \emph{XMM-Newton} observations of Mrk\,766;
\emph{bottom left} \emph{XMM-Newton} observations of 1H0707--495 \cite{miller10b}.
Lags are positive when the hard band lags the soft band.
\label{fig2}
}}
\end{minipage}
}
\end{center}

\end{figure}

Unlike the case of optical reverberation studies, it is not possible to isolate
a time series of purely emission-line fluxes, because on the required rapid variability
timescales, there are insufficient counts to obtain line flux values that are not
completely dominated by shot noise.  As in the case of optical studies, it also
seems likely that a significant component of line emission arises from distances
much larger than can be probed by the timescales in the X-ray observations.

Instead, however, we note that any circumnuclear material will Compton-scatter
the incident flux, and because of the strong energy dependence of the absorption
opacity of moderately-ionised material, we expect high energy photons to be much
more efficiently scattered than low energy photons: the latter are mostly absorbed.
Thus we can search for reverberation by cross-correlating hard and soft X-ray bands,
making them sufficiently broad in energy that shot noise is low, and we should expect
to see time delays between the hard and soft bands.  In fact, such time delays have
been measured for some years already in both AGN and Galactic binary systems,
e.g. \cite{papadakis,mchardy04,mchardy07},
but their significance as an indicator of reverberation has not previously
been recognised, and the lags have instead been interpreted by a model of emission from
radially propagating fluctuations in an accretion disk \cite{arevalo06}.

Here, we have analysed three AGN that are also NLS1, from
\emph{Suzaku} observations totalling 473\,ks of NGC\,4051 \cite{miller10a} and
\emph{XMM-Newton} observations totalling 493\,ks of Mrk\,766 \cite{miller07} and 
452\,ks of 1H\,0707--495 
\cite{miller10b}.  
Fig.\,\ref{fig2} shows clear transitions from positive lags (hard band lagging
soft) at low frequencies to zero lags at high frequencies.  The transitions are
not as sharp as expected for thin shells of material, indicating that the reverberating
material has some depth \cite{miller10a}.  The light travel times across the regions
in each case are approximately 10\,ks, 5000\,s and 2000\,s respectively, indicating
the detection of reverberating material a few light hours from the central source.
Some simple models of the lag spectra are presented by \cite{miller10a, miller10b}.

\section{1H\,0707--495}
One noticeable feature of the lag spectrum of 1H\,0707--495 is that the lags
are significantly negative at high frequency \cite{fabian09, zoghbi10}: i.e.
the soft band lags the hard band, contrary to the expectation outlined above.
This has been interpreted as evidence for extreme inner-disk reflection from
the accretion disk.  A model describing such lags ascribes the lagging of the soft band 
to two effects, both of which are required: 
(i) a factor $> 9$ overabundance of Fe; and (ii) amplification
of the Fe disk reflection line by the process of ``light-bending'', in which
light rays near the black hole are preferentially bent onto the accretion disk,
which then produces stronger-than-expected reflection \cite{miniutti}.  Thus, it
is claimed, extremely strong reflection Fe\,L lines may be produced in the
$0.4-1$\,keV band, which lead to the apparently inverted lag spectrum. Fe\,L
lines are not visible in the spectrum, and the spectral model supposes that the
lines are highly relativistically-blurred (a model-dependent view of these line
components has been shown by \cite{fabian09}).

\begin{figure}
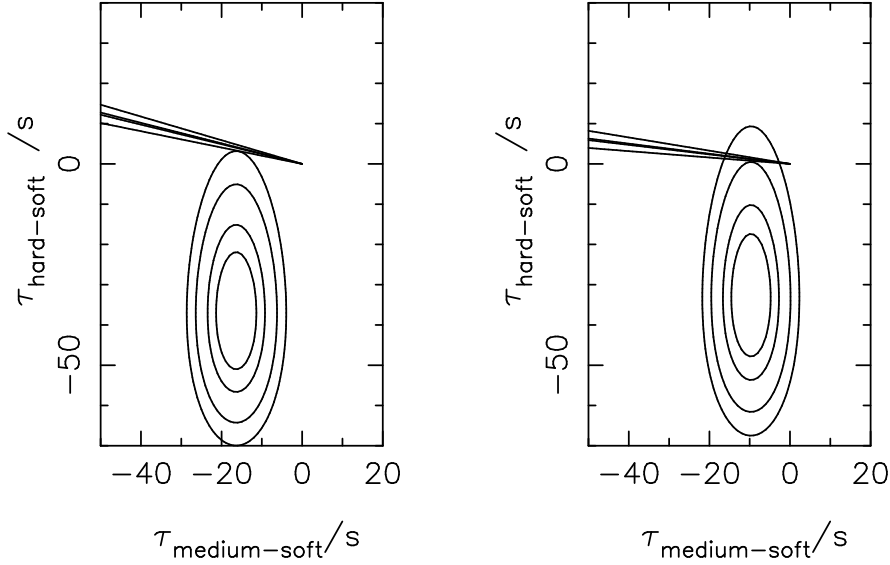

\begin{center}
\resizebox{0.35\textwidth}{!}{
\rotatebox{-90}{
\includegraphics{hardsoftchisq_original.ps}
}}
\hspace*{1cm}
\resizebox{0.35\textwidth}{!}{
\rotatebox{-90}{
\includegraphics{hardsoftchisq.ps}
}}
\end{center}
\caption{
Confidence regions for the time delays between the hard and soft 
bands (y-axis) \emph{v.} the time delays between the medium and
soft bands (x-axis) in 1H\,0707--495 
at high frequencies ($0.00066 < \nu < 0.0053$\,Hz).
Contours show the 50, 90, 99 and 99.9\,percent confidence regions, to be compared
with the inner-disk reflection spectral models - straight lines
with a range of slopes indicate the model expectation and the modelling
uncertainty (see \cite{miller10b} for full details):
\emph{left} the original model of Zoghbi et al. (2010) \cite{zoghbi10}; 
\emph{right} the revised model \cite{zoghbi11},
which includes a reverberating black body.  
In the left panel the band definitions are
0.4--1\,keV (soft), 1--4\,keV (medium) and 4--7.5\,keV (hard) (see \cite{miller10b}):
in the right panel the soft band has been modified to be 0.5--1\,keV 
(see text for details).  
The original inner-disk reflection model is ruled out at 99.9\,percent confidence,
the revised model is strongly disfavoured at 99\,percent confidence.
\label{fig3}
}
\end{figure}

However, it has been shown above how lag spectra may easily become negative in
some ranges of frequency, even if the hard band always is lagged more than the 
soft band (see also \cite{miller10b}).  The claim of inner-disk reflection 
ignores the strong positive lags, up to 600\,s, that are clearly seen in Fig.\,\ref{fig2}
and only addresses the lags that are negative by a few tens of seconds.  
There are substantial difficulties with this explanation:
\begin{enumerate}
\setlength{\itemsep}{-1ex}
\item The proposed inner-disk reflection requires two independent 
mechanisms for generating
lags: negative ones at high frequency and positive ones at low frequency.
The proposed model requires some fine tuning to allow the
inner disk reflection to dominate at high frequencies, but the positive-lag mechanism
to dominate at low frequency.
\item The model requires light-bending to produce sufficient soft-band line emission,
which in turn requires the existence of a hypothetical compact source, close to
the black hole, but a small distance above the accretion disk.  If the positive lags
arise from radially-propagating fluctuations in the accretion disk itself, these
two mechanisms are mutually exclusive.  Either the emission comes from the 
light-bending ``lamp-post'', or it comes from the accretion disk: it cannot arise
from both.  Such a model could only be tenable if there were some
mechanism by which emission from the lamp-post were excited by the accretion
disk fluctuations, without those fluctuations themselves producing emission.
\item The proposed distance between source and reflector is $\sim 1$\,r$_g$
for a black hole of mass $2\times 10^6$\,M$_\odot$.
However, if radiating at the Eddington limit, the black hole's mass is likely to
be about a factor 10 higher \cite{leighly}, so the separation between source and
reflector would be $\sim 0.1$\,r$_g$, inconsistent with the light-bending model.
\item There is no independent evidence for the highly super-solar Fe abundance that
is also required.
\item The analysis of \cite{fabian09,zoghbi10} only tested for lags between
the soft band ($0.4-1$\,keV)
and the medium band ($1-4$\,keV).  However, if there is substantial Fe\,L
line emission in the soft band, there should be even more line emission in the
Fe\,K band.  This was tested by Miller et al. \cite{miller10b} who found that
the spectral model of \cite{zoghbi10} indeed predicted that there should be more
hard band reflection than in any other band, and therefore that the lags between
the hard band and either the soft or medium bands should be positive, as usually 
expected when the hard band dominates the reflection.  It was shown that the lags
between these bands in fact remain negative at high frequencies (see Fig.\,\ref{fig2}), 
and the inner-disk
reflection model was ruled out at 99.9\,percent confidence \cite{miller10b} 
(Fig.\,\ref{fig3}).
\end{enumerate}
Notwithstanding the above difficulties, a revised inner-disk reflection model
has been presented, in which a component of soft-band black body emission previously
included is now also supposed to have reverberation time delay \cite{zoghbi11}.  
This leads to further difficulties for the model:
\begin{enumerate}
\setlength{\itemsep}{-1ex}
\setcounter{enumi}{5}
\item Normally it is supposed that thermal radiation from the accretion disk is
upscattered to the X-ray regime, and the thermal emission should therefore
\emph{lead} the non-thermal emission in its variations.  In the proposed model, the
thermally-emitting material does not intrinsically vary, but reflection from
its outer layers responds to the varying non-thermal source (invisible
propagating fluctuations in the accretion disk are still required to produce the
positive lags at low frequencies).  The spectrum of such 
reflection would not be a black body and should be modelled following \cite{ross}.
It should also be relativistically-blurred. Neither effect has been included.
Furthermore, any reflection variations would be substantially diluted by the
thermal emission \cite{ross}.
\item The proposed thermal component only produces significant emission below
0.5\,keV, so we may readily repeat the test shown in Fig.\,\ref{fig3} excluding
the regime below 0.5\,keV.  We find that although the confidence intervals
are broader because of the loss of data, the inner-disk reflection model is
still excluded at confidence level 99\,percent, even though this model was
designed to circumvent this difficulty.
\end{enumerate}
Our conclusion is that the inner-disk reflection model is not viable.
In principle, the negative lags may be produced as an artifact of the Fourier
filtering shown in Fig.\,\ref{fig1}: however, it is difficult to obtain lags that
remain negative over a broad range of frequencies.  An alternative explanation
is that all the bands being cross-correlated contain both direct and scattered
components, with differing transfer functions in each band.  Simple top hat transfer
function models
can reproduce the observed lag spectrum between the various energy bands \cite{miller10b}.
We should also take account of the finite size of the X-ray emission region: 
this is likely to be a Compton upscattering corona with size of order ks and with
intrinsic time delays of order tens of seconds (e.g. \cite{kazanas}).

\section{Conclusions}
X-ray observations of highly-variable NLS1 AGN reveal time delays between variations
at hard photon energies with respect to soft photon energies.  The time delays
have values of hundreds of seconds at low variation frequencies but fall to values
close to zero at high frequencies.  The transition frequencies occur at $\sim 10^{-4}$Hz.
Such a lag spectrum signature is expected in models of reverberation, where the 
transition occurs at a time period comparable to the light crossing time of the
reverberating region.  This indicates the existence of scattering material at
distances of a few light hours from the central source, equivalent to a fews tens
to a few hundreds of gravitational radii.   Simple reverberation models have been
presented elsewhere \cite{miller10a, miller10b}.  It seems likely that the global covering
factor of the scattering material must be high ($\sim 50$\,percent) in order to
achieve a sufficient amount of scattered light.  

One NLS1, 1H\,0707--495, displays soft band lags at high frequencies, which we interpret
as being due to scattered light present at all photon energies in this source.
An alternative model ascribing the soft-band lags to inner disk reflection is 
strongly disfavoured by consideration of the full energy dependence.  

The large-scale reverberation explanation is supported by spectral evidence that
there are substantial zones of partially-covering, absorbing circumnuclear material
\cite{turnermiller}.  Further work is needed to create models that reproduce both
the timing behaviour and the observed high resolution spectra (e.g. \cite{sim08, sim10}).

\end{document}